# Status and prospect for determining $f_B$, $f_{B_s}$, $f_{B_s}/f_B$ on the lattice[1]

The MILC Collaboration

C. Bernard,[a] T. Blum,[b] A. De,[a] T. DeGrand,[c] C. DeTar,[d] Steven Gottlieb,[e]
Urs M. Heller,[f] N. Ishizuka,[a] L. Kärkkäinen,[b] J. Labrenz,[g] K. Rummukainen,[e]
A. Soni,[h] R. Sugar,[i] and D. Toussaint[b]

[a]*Department of Physics, Washington University,
St. Louis, MO 63130, USA*

[b]*Department of Physics, University of Arizona,
Tucson, AZ 85721, USA*

[c]*Physics Department, University of Colorado,
Boulder, CO 80309, USA*

[d]*Department of Physics, University of Utah,
Salt Lake City, UT 84112, USA*

[e]*Department of Physics, Indiana University,
Bloomington, IN 47405, USA*

[f]*SCRI, The Florida State University,
Tallahassee, FL 32306-4052, USA*

[g]*Physics Department, University of Washington,
Seattle, WA 98195, USA*

[h]*Physics Department, Brookhaven National Laboratory,
Upton, NY 11973, USA*

[i]*Department of Physics, University of California,
Santa Barbara, CA 93106, USA*

---

[1]Talk presented by Urs M. Heller at LISHEP95, February 20–22, Rio de Janeiro, Brazil.

**Abstract**

Preliminary results from the MILC collaboration for $f_B$, $f_{B_s}$, $f_D$, $f_{D_s}$ and their ratios are presented. We compute in the quenched approximation at $\beta = 6.3$, 6.0 and 5.7 with Wilson light quarks and static and Wilson heavy quarks. We attempt to quantify all systematic errors other than quenching, and have a first indication of the size of quenching errors.

# 1  Introduction

Existing and planned experimental measurements of $B$-$\bar{B}$ and $B_s$-$\bar{B}_s$ mixing do not constrain the Cabibbo-Kobayashi-Maskawa matrix without knowledge of the heavy-light decay constants $f_B$ and $f_{B_s}$ and the corresponding $B$-parameters. For example, the amplitude of $B$-$\bar{B}$ mixing is written in the form

$$x_d \sim |V_{td}^\star V_{tb}|^2 f(m_t) \cdot \frac{8}{3} m_B^2 f_B^2 B_B \ . \tag{1}$$

Here the mixing matrix element is

$$\frac{8}{3} m_B^2 f_B^2 B_B = \langle B_0 | (\bar{d}\gamma_\mu(1-\gamma_5)b)(\bar{d}\gamma_\mu(1-\gamma_5)b) | \bar{B}_0 \rangle \tag{2}$$

and the decay constant $f_B$ is defined by the matrix element

$$\langle 0 | \bar{d}\gamma_\mu\gamma_5 b | B(p) \rangle = ip_\mu f_B \ . \tag{3}$$

The only known method of computing $f_B$ and the $B$-parameter $B_B$ from first principles is lattice QCD. This fact has led to a major effort in the lattice community to compute these quantities [1].

The $B$-parameter is believed to be close to its vacuum saturation value $B_B \approx 1$ and the major uncertainty is thus the value of $f_B$. Fortunately this is also the easier matrix element to be computed with lattice methods, since it only involves the computation of appropriate two-point functions and not the harder three-point functions needed for the computation of $B_B$.

Over the past year and a half, the MILC collaboration has been computing heavy-light decay constants in the quenched approximation – the approximation where internal fermion loops are ignored – on Intel Paragon computers. Most of the computations have been performed on the 512-node Paragon at Oak Ridge National Laboratory, but Paragons at Indiana University and at the San Diego Supercomputer Center have also been used. More recently we have started computations with the effects of two light flavors of dynamical fermions included. This will allow us, eventually, to remove the last unknown systematic uncertainty in our computations.

## 2 Features of the lattice computation

Lattice QCD is a discrete approximation to QCD where the fields are restricted to the sites (quark fields) and the links connecting neighboring sites (gauge fields) of a space-time lattice with lattice spacing $a$. Continuum physics is recovered in the limit $a \to 0$ and thus all results have to be carefully extrapolated to this limit. Next, the space-time volume is kept finite and the results have thus to be extrapolated to the infinite volume limit. Keeping both lattice spacing and volume finite we have a finite number of degrees of freedom which allows us to use powerful numerical methods for our computations. We use stochastic Monte Carlo methods and therefore our results will have statistical errors.

To calculate $f_B$, we compute, with $\hat{B}$ an operator that creates a B-meson, *e.g.*, $\hat{B} = \bar{d}\gamma_5 b$, the two-point function

$$\sum_{\vec{x}} \langle (\bar{d}\gamma_0\gamma_5 b)(\vec{x},t)\hat{B}^\dagger(0)\rangle = \sum_n \langle 0|\bar{d}\gamma_0\gamma_5 b|n\rangle \frac{e^{-E_n t}}{2E_n V}\langle n|\hat{B}^\dagger|0\rangle$$

$$\xrightarrow[t\to\infty]{} \frac{m_B f_B}{2m_B V}\langle B|\hat{B}^\dagger|0\rangle e^{-m_B t} \; . \qquad (4)$$

The unknown amplitude $\langle B|\hat{B}^\dagger|0\rangle$ can be obtained from the two-point function

$$\langle \hat{B}\hat{B}^\dagger\rangle \xrightarrow[t\to\infty]{} \frac{1}{2m_B V}|\langle B|B^\dagger|0\rangle|^2 e^{-m_B t} \; . \qquad (5)$$

The parameters in eqs. (4,5) are obtained by fits to the (stochastically) computed two-point functions.

Performing such a computation we still do not yet get quite the desired result. Though the current we have used for our computation on the lattice looks identical to the continuum current, quantum fluctuations induce a finite renormalization

$$f_B^{cont} = Z_A^{-1} f_B^L, \qquad Z_A = 1 + \mathcal{O}(\alpha_s), \qquad (6)$$

computable in perturbation theory. All our results we will quote include this renormalization.

For the computation of the two-point functions we need one light and one heavy quark propagator. We compute the heavy quark propagator in a "hopping parameter" expansion [2] – essentially a $1/m_Q$ expansion – keeping 400 terms. This gives very good convergence, on our lattices, for quark masses well below the charm. The ability to adjust the heavy quark mass arbitrarily is proving very useful in the analysis of systematic errors.

Since we only have results for degenerate light quarks, we determine $\kappa_s$, the strange quark hopping parameter, by adjusting the pseudoscalar mass to $\sqrt{2m_K^2 - m_\pi^2}$, the lowest order chiral perturbation theory value.

For heavy-light mesons we use the Kronfeld-Mackenzie [3] norm ($\sqrt{1-6\tilde{\kappa}}$) and adjust the measured meson pole mass upward by the difference between the heavy quark pole mass ("$m_1$") and the heavy quark dynamical mass ("$m_2$") as calculated in the tadpole-improved tree approximation [3]. This procedure eliminates the leading (in $\alpha_s$) discretization errors that would have been of order $\mathcal{O}(am_Q)$, with $am_Q \gtrsim 1$ in the charm region.

# 3 Results

Table 1: Lattice parameters.

| name | $\beta$ | size | # configs. |
|------|---------|------|------------|
| A | 5.7 | $8^3 \times 48$ | 200 |
| B | 5.7 | $16^3 \times 48$ | 100 |
| E | 5.85 | $12^3 \times 48$ | 100 |
| C | 6.0 | $16^3 \times 48$ | 100 |
| D | 6.3 | $24^3 \times 80$ | 100 |
| F | 5.7; m=.01 ($n_F = 2$ QCD) | $16^3 \times 32$ | 49 |

So far we have results from six different lattices with parameters listed in Table 1. $\beta = 6/g^2$ denotes the bare lattice coupling.

In the limit of an infinitely heavy quark, heavy quark effective theory tells us that the combination $\phi_P = f_P\sqrt{M_P}$ of pseudoscalar meson decay constant and mass becomes independent of the heavy quark mass, up to renormalization group logarithms, with corrections that vanish as powers of $1/M_P$

$$\phi_P = f_P\sqrt{M_P} = \phi_\infty \left[1 + \frac{c_1}{M_P} + \frac{c_2}{M_P^2} + \cdots\right]. \tag{7}$$

Therefore, it has become customary to plot $f_P\sqrt{M_P}$ vs. $1/M_P$. Such a plot for lattice D is shown in Fig. 1. The fit is covariant, to the form (7). Although to the eye there appears to be reasonable consistency among the heavy-light results and between the heavy-light and static-light results, the $\chi^2$/d.o.f for the fit is $\approx 2$ (confidence level $\approx 10\%$), whether or not the static-light point is included. The rather low confidence level may perhaps be due to the fact that we have not included additional large-$ma$ corrections to the action and operators [4], or simply to the small differences between the heavy quark mass and the meson mass $M_P$. Such effects are under investigation. Note that, in an earlier calculation [5], the statistical errors were considerably larger, and the $\chi^2$/d.o.f for such fits was good. Here the statistical precision has increased to a level where small effects are becoming important.

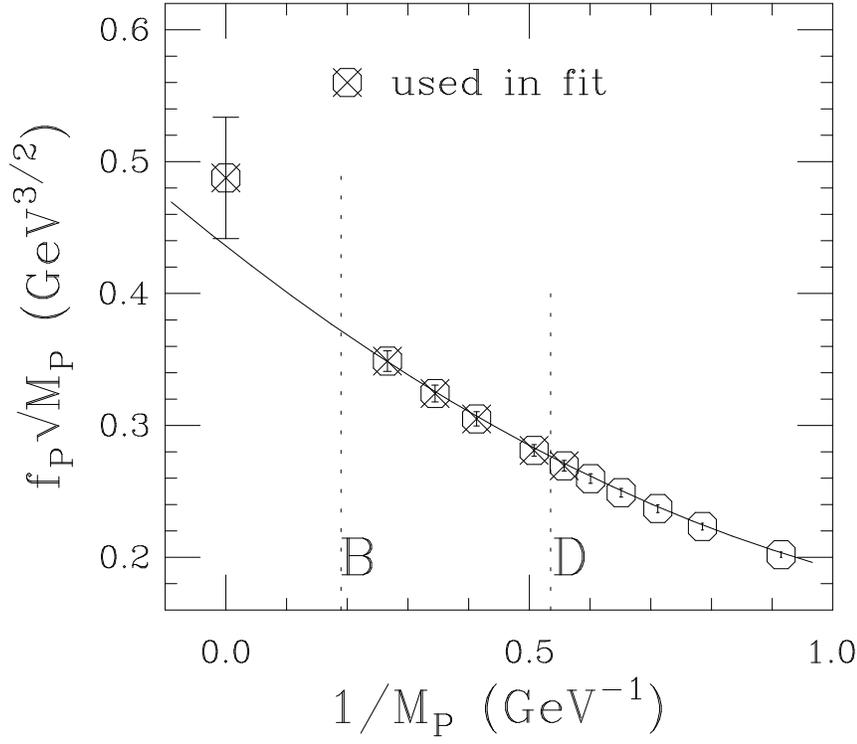

Figure 1: $f_P(M_P)^{\frac{1}{2}}$ vs. $1/M_P$ for lattice D. The light quark is extrapolated to the physical mass $(m_u + m_d)/2$.

Table 2 shows results from the six lattices. The lattice-spacing dependence is appar-ent, but little, if any, finite volume effect is present (compare A and B). This is seen more clearly in Fig. 2, which shows $f_B$ vs. lattice spacing. It is natural to extrapolate the quenched $f_\pi$-scale results linearly to the continuum; we get 148(5) MeV. Note that the $f_\pi$-scale results have much less $a$ dependence than those using $m_\rho$. This makes sense since $f_\pi$ and $f_B$ are likely to have rather similar finite $a$ effects. Neither a linear nor a linear + quadratic fit to the $m_\rho$-scale results is particularly good; the linear + quadratic fit is somewhat better and gives a result which is consistent, within large statistical errors, with the $f_\pi$-scale results: 158(25) MeV.

Table 2: Results for decay constants and ratios. $f_\pi$=131 MeV scale used throughout.

|  | A | B | E | C | D | F |
|---|---|---|---|---|---|---|
| $f_B$ | 195(6) | 198(4) | 190(4) | 176(3) | 166(4) | 195(4) |
| $f_{B_s}$ | 244(5) | 237(3) | 226(3) | 206(3) | 192(3) | 220(4) |
| $f_D$ | 227(5) | 227(4) | 216(3) | 203(2) | 198(2) | 219(2) |
| $f_{D_s}$ | 275(4) | 273(3) | 253(3) | 236(2) | 225(2) | 242(2) |
| $\frac{f_{B_s}}{f_B}$ | 1.25(2) | 1.20(1) | 1.19(1) | 1.17(1) | 1.16(1) | 1.13(1) |
| $\frac{f_{D_s}}{f_D}$ | 1.21(1) | 1.20(1) | 1.17(1) | 1.16(1) | 1.13(1) | 1.11(0) |

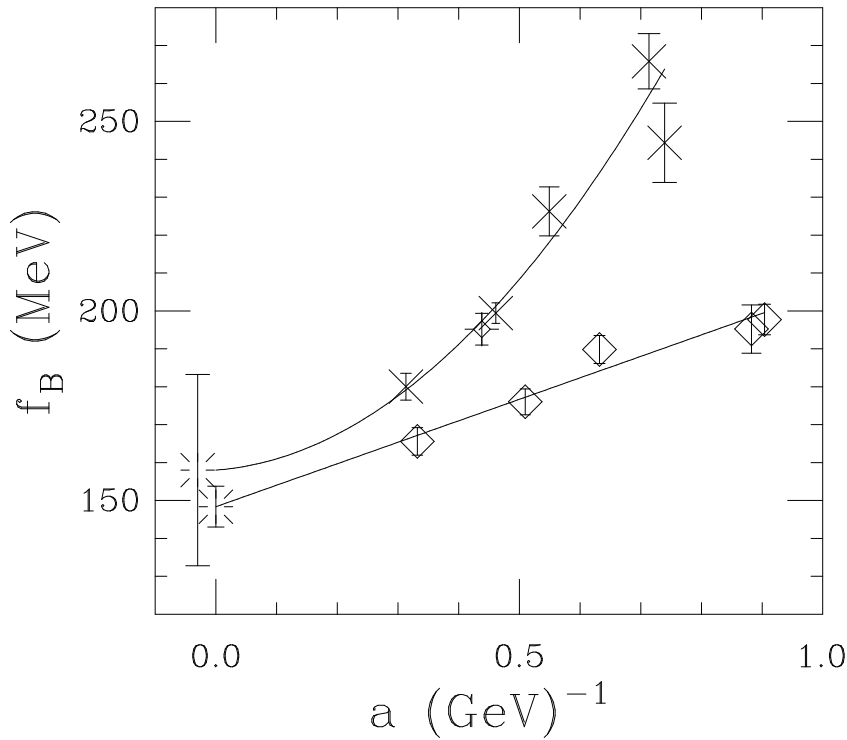

Figure 2: $f_B$ vs. $a$. Diamonds have scale set by $f_\pi$; crosses, by $m_\rho$. Fits are to the diamonds (linear) and to the crosses (linear + quadratic); the $a = 0$ extrapolations are marked by "bursts." The higher cross at $a \approx 0.7$ is from lattice B. The "fancy diamond" is from configuration F, $n_F = 2$ "full QCD," $f_\pi$ scale.

The results on lattice D are consistent with those of [5] which uses the same $\beta$ and volume. The major cause of the somewhat smaller central values here is a lower (but still consistent) estimate of the scale ($1/a \approx 3.0$ GeV here vs. $\approx 3.2$ GeV in [5]). The difference is within the statistical scale error quoted in [5].

We linearly extrapolate to $a=0$ all results in Table 2. Systematic errors are then estimated — in a preliminary fashion — as follows:

- Changes of fitting ranges (in $t$) for the propagators and of types of fits in $1/M$ for $f_P\sqrt{M_P}$ give a typical variation of about twice the statistical errors.

- The dependence on the determination of $\kappa_s$ is estimated by finding the change in the extrapolated results if $\kappa_s$ is fixed using the vector state $\phi$ instead of the pseudoscalar state. The difference is especially significant for $f_{B_s}/f_B$ and is $\approx 0.06$ there.

- Finite volume effects are estimated by taking the fractional difference between results from lattices A and B, using the $f_\pi$ scale.

- We estimate scale errors by comparing the extrapolated results from the $f_\pi$ and $m_\rho$ scales. The difference ($\approx 10$ MeV for the decay constants and $\approx 0.02$ for the ratios) is comparable to (but slightly less than) what we would get by comparing $f_\pi$ $m_\rho$-scale values at the smallest available lattice spacing ($\beta = 6.3$).

- The effects of using heavy Wilson fermions without the additional corrections to the action and operators detailed in [3, 4] are estimated by comparing the original fits (see, e.g., Fig. 1) with fits using only the 5 or 6 lightest heavy-light states (and, where appropriate, the static-light point). At $\beta = 6.3$, the maximum value of $(m_2 - m_1)/m_2$ is 0.22 with the original fits and 0.04 with the new ones. The differences in the results are quite small: $\approx 4$ MeV for the decay constants and $\approx 0.01$ for the ratios.

- The effects of quenching are estimated by comparing the results of $n_F = 2$ simulations with the quenched results interpolated to the same lattice spacing. (See for instance Fig. 2 for the case of $f_B$.) Note that this seems likely to give an overestimate of the quenching effects, since quenched chiral perturbation theory arguments [6] indicate that "real" ($n_F = 3$) QCD is closer to the quenched theory than $n_F = 2$ QCD. The estimated uncertainty from quenching is of order 10%, i.e., about 3-4 times the statistical errors, for the decay constants, and of order 5% for the ratios of the decay constants. (This is reasonable since some cancellations are expected in the ratios.)

Adding all the above systematic errors except quenching in quadrature, our preliminary results are

$$f_B = 148(5)(14)(19), \quad f_D = 180(4)(18)(16), \tag{8}$$
$$f_{B_s} = 165(4)(20)(17), \quad f_{D_s} = 194(3)(16)(9), \tag{9}$$
$$\frac{f_{B_s}}{f_B} = 1.13(2)(9)(4), \quad \frac{f_{D_s}}{f_D} = 1.09(1)(4)(4), \tag{10}$$

where the decay constants are in MeV. The first error is statistical, the second is due to non-quenching systematics, and the third is an estimate of the quenching effect. The systematic error estimates on $f_B$ and $f_{B_s}/f_B$ have been recently updated. Further study of the systematic errors, and especially the quenching error, is in progress.

## 4 Future prospects

We see that the errors on the heavy decay constants and their ratios are dominated by the systematic errors, which are typically a factor of 4–5 larger than the statistical errors. Therefore, progress will come from reducing the systematic errors. In particular we foresee progress by

- Removing O(a) effects by using improved fermionic actions (clover fermions); a more ambitious step, correcting for all orders in $ma$ up through order $\alpha_s$, is being attempted by the Fermilab group [4].

- Diminishing O(a) effects by using weaker coupling and hence smaller lattice spacings: work on a simulation with $\beta = 6.52$ on a $32^3 \times 100$ lattice has begun.

- More unquenched results. Since quenching is in many cases the dominant uncertainty this appears to be the most fruitful venue. Unfortunately it is also the hardest, requiring by far the most computational resources. Existing unquenched configurations have relatively large lattice spacings and small volumes, but with the advent of more and more powerful computers progress will come.

We expect that in about two years the systematic errors will be reduced to $\pm 10$ MeV in $f_B$ (not including quenching) with an error of $\sim \pm 12\%$ from quenching. Similarly, we expect systematic errors of $\pm .04$ in $f_{B_s}/f_B$ (not including quenching) plus an error of $\sim \pm .03$ from quenching. Since the results seem to depend sensitively on $n_F$, the prospect of correcting for quenching (as opposed to merely estimating the error) is further away: one will need simulations with $n_F = 3$ flavors of quarks, two light flavors for the $u$ and $d$ quarks, and one somewhat heavier, for the $s$ quark.

We are grateful to Norman Christ and the Columbia group for providing the $n_F = 2$ QCD configurations, and to Greg Kilcup for help in reading and transferring them. We thank A. El-Khadra, A. Kronfeld, and P. Mackenzie for useful conversations. Computing was done at ORNL Center for Computational Sciences, SDSC, and Indiana University. This work was supported in part by the DOE and NSF.